\documentstyle[rotate]{l-aa}
\def\sun{\hbox{$\odot$}}

\begin{document}
\input{psfig.sty}

\thesaurus{}

\title{NGC\,1427A -- an LMC type galaxy in the Fornax Cluster}

\author {Michael Hilker \inst{1,2}, Dominik J. Bomans \inst{3,4}, Leopoldo 
Infante \inst{2} \& Markus Kissler-Patig \inst{1,5}
} 

\offprints{Michael Hilker}

\institute{
Sternwarte der Universit\"at Bonn, Auf dem H\"ugel 71, 53121 Bonn, Germany;
email: mhilker@astro.uni-bonn.de
\and
Departamento de Astronom\'ia y Astrof\'isica, P.~Universidad Cat\'olica,
Casilla 104, Santiago 22, Chile
\and
University of Illinois at Urbana-Champaign, 1002 West Green Street, 
Urbana, IL 61801, USA;  email: bomans@astro.uiuc.edu
\and
Feodor Lynen-Fellow of the Alexander von Humboldt-Gesellschaft
\and
Lick Observatory, University of California, Santa Cruz, CA 95064, USA 
}

\date {}

\maketitle
\markboth{Hilker et al.: NGC\,1427A -- an LMC type galaxy in the Fornax Cluster}
{Hilker et al.: NGC\,1427A -- a LMC type galaxy in the Fornax Cluster}

\begin{abstract}

We have discovered that the Fornax irregular galaxy NGC\,1427A is in very many
respects a twin of the Large Magellanic Cloud. Based on $B$, $V$, $I$, and 
H$\alpha$ images, we find the following.
The light of the galaxy is dominated by high surface brightness regions in
the south-west that are superimposed by a half-ring of OB associations
and H\,{\sc ii} regions indicating recent star formation. 
The colors of the main stellar body are $(V-I) = 0.8$ mag and $(B-V) = 0.4$ mag,
comparable to the LMC colors.
A low surface brightness cloud north of the main body as well as a LSB tail in 
the west have colors ($V-I \simeq 0.2$ mag) that are more typical for blue
compact dwarf galaxies.
We identified a system of cluster candidates with mean
ages $\leq$ 2 Gyr (assuming a LMC metallicity) comprising the richest cluster
system in an irregular galaxy observed up to now outside the Local Group.
In X-ray wavelengths NGC\,1427A appears with a relatively soft and complex 
spectrum.


\end{abstract}


\section{Introduction}

NGC\,1427A is the brightest irregular galaxy in the Fornax cluster.
It is about 3 magnitudes brighter than other Fornax galaxies that are 
classified as irregulars (Ferguson 1989).
Its total $B$ magnitude is $13.3\pm0.21$ mag (de Vaucouleurs et al. 1991).  
Assuming a distance of 16.4 Mpc (Kohle et al. 1996) to the Fornax cluster, 
the apparent $B$ magnitude corresponds to an absolute magnitude of 
$M_B = -17.8$ mag, which is nearly identical to the LMC luminosity 
$M_B = -17.93$ mag, (de Vaucouleurs et al. 1991), assuming a LMC distance 
modulus of $(m-M) = 18.5$ mag.
The linear dimension of NGC\,1427A is about 11 kpc in major axis ($\simeq
2\farcm4\times2\farcm0$ within the isophote at $\mu_V = 24.7$ mag/arcsec$^2$), 
slightly larger than the LMC which 
has a major axis diameter of $\simeq 9.4$ kpc. Table 1 summarizes some 
properties of NGC\,1427A compared with the LMC. The values are from the
Third Reference Catalogue of Bright Galaxies (de Vaucouleurs et al. 1991),
if no other reference is given.

The projected distance to the central Fornax galaxy NGC\,1399 is $22\farcm9$,
about 109 kpc at the cluster distance. Since NGC\,1399 has an extended cD halo
of about 500 kpc diameter (Schombert 1986, Killeen \& Bicknell 1988), one 
might expect tidal interaction of both galaxies.
However, the radial velocities, measured in HI (Bureau et al. 1996), of NGC\,
1427A ($v_r = 2023\pm10~{\rm km~s}^{-1}$) and NGC\,1399 ($v_r = 
1430\pm9~{\rm km~s}^{-1}$) differ by about
$600~{\rm km~s}^{-1}$ which is nearly twice the cluster velocity dispersion 
($\sigma_v = 325~{\rm km~s}^{-1}$, Ferguson \& Sandage 1990).
NGC\,1427A rather agrees with the velocity of NGC\,1404 ($v_r = 1944\pm15~
{\rm km~s}^{-1}$).  
NGC\,1404 is nearly as luminous as 
NGC\,1399 and has a projected distance to NGC\,1427A of about 75 kpc.

In this paper we discuss the similarity of the morphological properties of
NGC\,1427A 
and the LMC, as well as the stellar content, such as OB associations and
young clusters,
X-ray properties 
and the evolutionary state.

The organization of the paper is as follows: In section 2 we describe 
observation and reduction of the data, section 3 deals with the 
large scale light distribution of NGC\,1427A, 
section 4 discusses the colors and H$\alpha$ properties of the 
resolved sources, section 5 discusses the properties of the cluster 
candidates. 
Section 6 gives a short discussion of the X-ray informations and
in section 7 we summarize our results.

\begin{table}
\caption{Some properties of NGC\,1427A and of the LMC}
\begin{minipage}{8.5cm}
\begin{tabular}{l c c}
\hline
Property & NGC\,1427A & LMC \\
\hline
distance adopted [kpc] & 16400 & 50.1\\
$B_t^0$ [mag]& 13.3 & 0.57 \\
$V_t^0$ [mag]& 12.9 & 0.13 \\
$(B-V)_0$ [mag]& 0.4 & 0.44 \\
$(V-I)_0$ [mag]& 0.8 & ? \\
$M_B$ [mag]& -17.8 & -17.93 \\
$M_V$ [mag]& -18.2 & -18.37 \\
major axis [kpc] & 11 & 9.5 \\
inclination\footnote{Bureau et al. 1996 
(inclination from face-on)}$^,$\footnote{Westerlund 1990
(inclination from face-on)} 
[$^\circ$] & $48\pm5$ & $\simeq 45$ \\
dist. to Milky Way [kpc] &  & 50 \\
proj. dist. to 1399 [kpc] & 109 & \\
proj. dist. to 1404 [kpc] & 75 & \\
$F_{\rm HI}$ [Jy km sec$^{-1}]^{\it a,}$\footnote{Huchtmeier et al. 1988} &
$23.1\pm 1.2$ & $(1.04\pm 0.12)\cdot 10^6$ \\
log (HI mass) [M$_{\sun}]$\footnote{formula of Roberts 1975} & 8.79 & 9.17 \\
\hline
\end{tabular}
\end{minipage}
\end{table}


\section{Observations and reductions}

We obtained long $V$ and $I$ exposures ($3 \times 900$~s in each color) with the 
100\arcmin~DuPont
telescope at Las Campanas Observatory, Chile, in the nights of 26-29 
September, 1994.  We used a Tektronix $2048\times 2048$ pixel chip, with a 
pixel size of $21\mu$m or $0\farcs227$ at the sky, corresponding to a total 
field of view of $7\farcm74\times7\farcm74$. The data reduction and 
calibration is described in Kissler-Patig et al. (1996).  
The seeing of $0.9\arcsec$, measured by the FWHM of stellar images,
corresponds to a linear diameter at the Fornax distance of about 72 parsec,
which sets our resolution limit (e.g. globular clusters would appear as 
point sources). 

The $B$ exposures ($3 \times 1200$~s) were obtained with the 40\arcmin~Swope 
telescope 
at Las Campanas Observatory, Chile, in the night of 6/7 December, 1996.
We used a SITe \#1 $2048\times 2048$ pixel chip, with a pixel scale of
$0\farcs694$ at the sky. The seeing was $1.5\arcsec$.

The $B$, $V$, and $I$ magnitudes of point sources inside 
NGC\,1427A and in its surroundings were derived in the following way:
first, we subtracted all objects from the images which were found and fitted
with Daophot II (Stetson 1987, 1992). On this cleaned image we replaced all
extended objects by values that were interpolated from the surrounding sky or
the unresolved stellar component of NGC\,1427A. We then smoothed these images 
by a $7 \times 7$ pixel median filter and finally subtracted the results
from the original images.
The final photometry was done with Daophot II on these images.
The colors of all objects were measured on the same
images by aperture
photometry assuming a constant background of the sky. We used a circular 
aperture of $2.0\arcsec$.

The H$\alpha$ and $R$ exposures were obtained at Cerro Tololo with the
0.9m telescope in the night of 1/2 December, 1995. We exposed $3 \times 1200$~s 
in H$\alpha$ and 600~s in $R$.
The seeing was about $1.0\arcsec$.
The calibration was done by using Landolt (1992) fields.

We retrieved HST WFPC2 images of NGC\,1427A in the F606W filter with 
exposure time of $2\times 80$~s from the STScI data archive. 
The effective resolution is $0.19\arcsec$, or about 15 pc in Fornax 
distance. Since the exposure is not deep, we used it only to check for 
multiplicity of the brightest point sources on the ground-based images
where we had overlap. 

X-ray observations of NGC\,1427A were retrieved from the High Energy 
Astrophysics Science Archive Research Center (HEASARC) at Goddard Space
Flight Center of NASA.  The ROSAT PSPC pointing RP600043 is centered on 
NGC\,1399. This pointing also includes NGC\,1427A inside the 
field of view.  An X-ray source is clearly present at the position of 
NGC\,1427A.  Unfortunately NGC\,1427A is located near the window support 
structure of the telescope, just outside the central 40\arcmin\ field of view.  
Therefore the 
spatial structure of this source is dominated by the point spread function and 
the rapid drop of sensitivity to the window support structure.  

\section{Large-Scale Morphology and Colors of NGC\,1427A}

The appearance of NGC\,1427A resembles that of the LMC with a bar-like main 
body and several regions of active star formation (Fig.~3).  
North of the `bar' a large blob of diffuse light and embedded point sources is 
located, detached from the main body of NGC\,1427A (in the following the
`nob' $=$ `northern blob'). This region appears 
similar to Shapley Constellation III north of the bar of 
the LMC. 

The appearance at low surface brightness shows that the main body
and the `nob' are embedded in a common envelope of unresolved stellar
light (see Fig.~4). The lowest isophote in Fig.~4 corresponds to a surface 
brightness of $\mu_V = 24.7$ mag/arcsec$^2$, which is 2$\sigma_{\rm sky}$ above 
the background. Inside this envelope, the most
luminous part of the main body lies off-centered to the south-west.
At the south-west border the regions with the highest surface brightness
are aligned in a half-ring. The most prominent of these regions is located at
(530,360) in (X,Y) pixel coordinates of Fig.~4. Appearence and size ($\sim
1.3$ kpc) of this star forming region strongly resembles the 30 Dor complex
in the LMC. The east side of the `bar' ends in an extended tail 
of low surface brightness ($\mu_V \simeq 23.0$ mag/arcsec$^2$). At the southern
border of the `nob' some low surface brightness spurs point towards the
main body of the galaxy (see Fig.~4).

\begin{figure}
\caption{This picture shows a $(V-I)$ color map of NGC\,1427A. The greyscale
corresponds to color bins of 0.1 mag as indicated in the plot. The sky is
arbitrarily set to a color of $(V-I) = 1.1$ mag. Arrows indicate the direction
to the central Fornax galaxies NGC 1399 and NGC 1404.}
\end{figure}
   
We constructed a calibrated color map from our $V$ and $I$ image to visualize 
the colors of the main body and the bright knots (Fig.~1).
All pixels with intensities lower than $2\sigma$ above the sky were arbitrarily
set to $(V-I) = 1.1$ mag. The color map was smoothed by a $7\times7$ pixel
median filter and the color values were binned in steps of 0.1 mag. The mean
color error is in the order of 0.03 mag in the region of the stellar body.
Several features become visible: the bright knots have colors bluer than
$(V-I) = 0.65$ mag. 
The stellar body has a mean color of $(V-I) = 0.8\pm 0.03$ mag except a 
north-south
stripe in the middle of the galaxy with colors around $(V-I) = 0.9\pm 0.03$ mag.
This redder color is probably due to a dust lane at the center of the
galaxy. Another explanation would be an older (redder) stellar population that
is dominating the color in this region.
The `nob' as well as the
faint luminosity tail in the east have significantly bluer colors of about 
$(V-I) = 0.7\pm 0.04$ mag, which indicates a different star formation history 
in these regions than in the main body.
 
We also measured $(B-V)$ colors of the stellar body in small
apertures assuming constant sky values. The mean colors are $(B-V) = 
0.4\pm 0.04$ mag
in the main body, which is comparable to the mean color of the LMC, $(B-V) = 
0.3\pm 0.04$ mag in the eastern tail, and $(B-V) = 0.2\pm 0.04$ mag in 
the `nob'.
Compared to other galaxies the `nob' lies in the color range of 
blue compact dwarf galaxies (hereafter BCDs) or very blue dwarf irregular 
galaxies (i.e. Thuan 1983, Gallagher \& Hunter 1987).

\begin{figure}
\psfig{figure=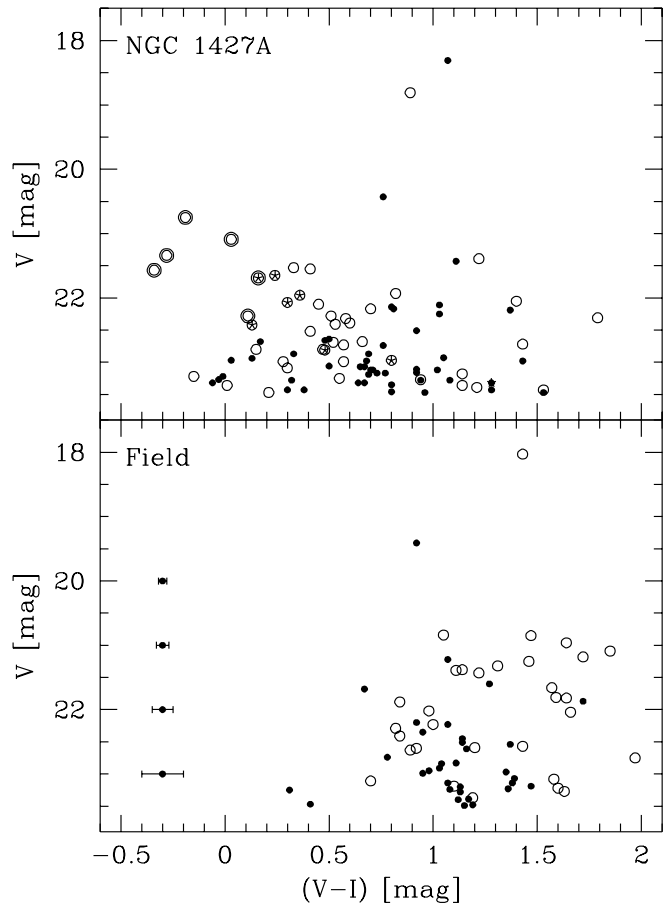,height=12.0cm,width=8.7cm
,bbllx=10mm,bblly=56mm,bburx=143mm,bbury=242mm}
\caption{The top panel shows a CMD of all objects within a radius of $1\farcm7$
from the center of NGC\,1427A. The dots are unresolved objects as discussed in
section 5. Circles indicate slightly extended objects
(the FWHM is
more than 1.5 times larger than for stellar images). Circles with stars inside
are objects with H$\alpha$ emission. Double circles turned out being
multiple objects after comparison with the HST image.
The lower panel shows the CMD of objects in a 2 times larger control
field north of NGC\,1427A. The mean errorbars of the color for different
magnitudes are indicated.}
\end{figure}

The blue color of the unresolved stellar light can not be explained by an old
stellar population alone, even if this population would have a very low 
metallicity of $Z = 0.001$. Therefore a young stellar component must have 
contributed to the light. We estimated ages by comparison with photometric 
evolutionary synthesis models of dI and BCD galaxies from 
Kr\"uger \& Fritze-v.~Alvensleben (1993).  In the case of continuous star 
formation a color of $(B-V) = 0.4$ of the main body is consistent 
with ages between 2 and 5 Gyr depending on metallicity and a constant or 
decreasing star formation rate (SFR). The stellar population of the 
`nob' with $(B-V) = 0.2$ would then be about 1 Gyr old.
Assuming a single starburst occuring in a 10 Gyr old galaxy with a previous
continuous SFR, colors bluer than $(B-V) = 0.4$ correspond to ages younger than
$10^8$ yr after the burst. 

\section{Extended Objects and H$\alpha$ distribution}

\begin{figure}
\caption{This picture shows a V image of NGC\,1427A. The white contours
indicate H$\alpha$ detections. The values are aperture $(V-I)$ and $(B-V)$
colors for the brightest, most prominent OB associations and clusters.
In comparison to Fig.~4 the (X,Y) pixel section of this image is (320,190)
for the lower left corner and (830,700) for the upper right corner.}
\end{figure}

The luminosities of point sources and slightly extended
objects in and around NGC\,1427A were determined by PSF fitting in the $B$, $V$,
and $I$ images. The colors were measured in apertures of $2.0\arcsec$ diameter
assuming constant sky values.
We present in the following foremost results using the $(V-I)$ colors,
because the resolution in the $B$ image is not as good as in the $V$ and $I$
images due to the larger pixel scale and the worse seeing. About 30\% of the
objects we found in $V$ and $I$ could not be measured in $B$.
Figure 2 shows in the upper panel the color magnitude diagram (CMD) of all
objects within a radius of about 8 kpc ($\simeq 1\farcm7$) from the 
geometrical center of NGC\,1427A (which is the center of the faintest 
isophote). In the lower panel the CMD for a two times larger control 
field located $5\arcmin$ north of NGC\,1427A is plotted. Objects redder than
$(V-I) = 0.8$ mag in both panels most probably are resolved as well as 
unresolved background galaxies and foreground stars. The slight excess of
extended objects in the control field compared to the target field can be 
explained by statistical fluctuations. 
Comparison of the two CMDs shows clearly that NGC\,1427A sample is 
dominated by very blue objects.

\begin{figure}
\caption{The location of the cluster candidates is given in the plot of
the isophotes of the main stellar body of NGC 1427A. The distribution
of the clusters is more uniform than that of the OB associations (see also
Fig.~3). The contours are derived from a $10\times10$ pixel median filtered
image .The 3 lowest isophotes correspond to $\mu_V = 24.7$ mag/arcsec$^2$,
$\mu_V = 24.1$ mag/arcsec$^2$, and $\mu_V = 23.6$ mag/arcsec$^2$, respectively.
Frame size and orientation are the same as in Fig.~1.}
\end{figure}

Figure 3 shows the V image of NGC\,1427A with H$\alpha$ contours overplotted.
Our H$\alpha$ image is not deep enough to search for diffuse filaments and 
shells similar the supergiant shell around Shapley Constellation 
III in the LMC (e.g. Meaburn 1980, Kennicutt et al. 
1995).  
The aperture $(V-I)$ and $(B-V)$ colors of the most prominent blue knots are 
indicated. 
Almost all regions that are associated with H$\alpha$ emission show very blue
components with colors in the range $-0.3 < (V-I) < 0.4$ mag (or $-0.1 < (B-V)
< 0.2$ mag). In some cases the $(V-I)$ color is negative, whereas the $(B-V)$ 
color is positive. This can only be explained by the strong contribution of
emission lines in the $V$ filter. The most prominent emission lines of 
H\,{\sc ii}
regions in the $V$ band are the O\,{\sc iii} lines at 4959 and 5007 \AA.
A comparison with the HST image shows that most of the bright blue knots are
multiple objects. Thus, from their color and sizes these objects can be
regarded as OB associations and H\,{\sc ii} regions.

\section{Properties of Cluster Candidates}

We defined as cluster candidates those objects that are clearly 
unresolved in our V image, i.e. those objects that have FWHM of the PSF smaller 
than $1.2\arcsec$ (see the dots in Fig.~2). The HST image shows that most 
of these
objects are also unresolved at its higher resolution, as one would expect,
since even the largest Milky Way globular clusters would appear hardly resolved
at Fornax distance.
(Note that the fainter clusters and objects in the planetary camera image could 
not be measured due to the short exposure time of the HST image).
We counted all objects within $80\arcsec$ radius from the geometrical 
center within the lowest isophotes. These objects are 
uniformly distributed over the whole area,
contrary to the light of the main stellar body and the OB 
association (see Fig.~4). 
 
\begin{figure}
\psfig{figure=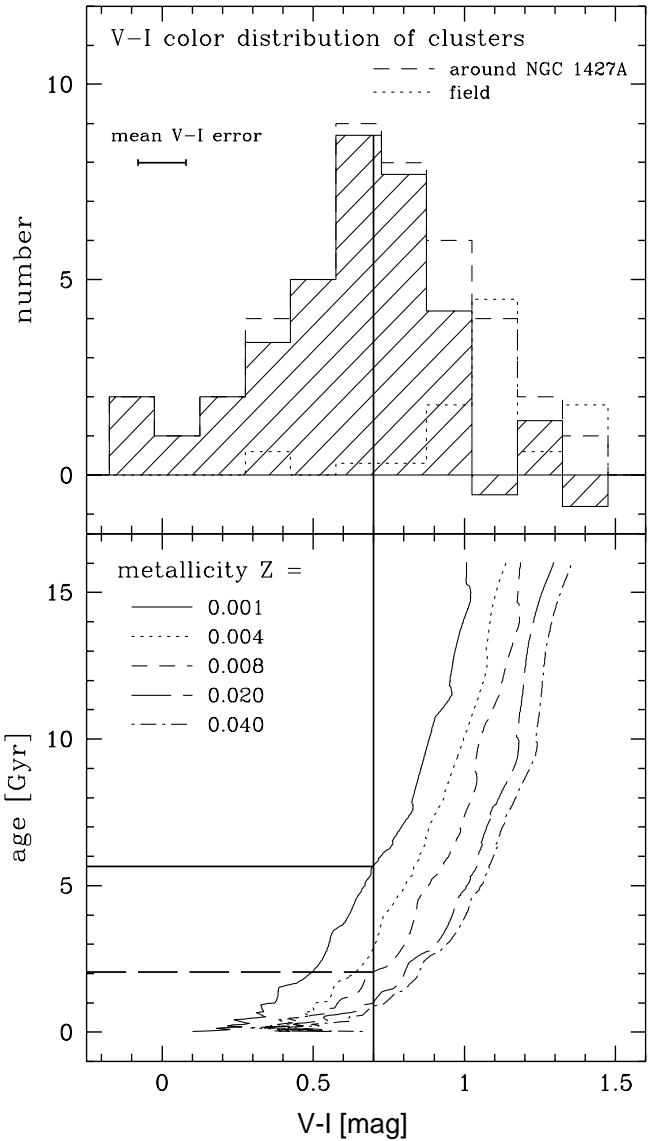,height=15.0cm,width=8.7cm
,bbllx=13mm,bblly=64mm,bburx=125mm,bbury=252mm}
\caption{The hashed histogram (upper panel) shows the $(V-I)$ color distribution
of clusters in NGC\,1427A after a statistical subtraction of background objects
(dotted histogram). The mean error is indicated. Note that the colors of the
bluest clusters are
contaminated by emission lines of nearby H\,{\sc ii} regions.
The lower panel illustrates the color evolution of clusters as function of
metallicity according to Fritze v.~Alvensleben \& Kurth (1997).
The peak color $(V-I)$ = 0.7 mag corresponds to an age of 
about 2.0 Gyr assuming an LMC metallicity.}
\end{figure}

Applying the same selection criteria as for the NGC\,1427A cluster
candidates we counted objects in a background control field $5\arcmin$ north of 
NGC\,1427A. We subtracted the area corrected background counts from the galaxy 
object counts in color bins of 0.15 mag as well as in magnitude bins of 0.5 mag
Figures 5, 6 (upper panels), and 7 show the resulting histograms.
The color histograms are selected for objects brighter than $V = 23.5$ mag.
As mentioned above not all objects could be measured in the B image.
Thus, the $(B-V)$ color distribution only contains about 80\% of the objects
of the $(V-I)$ color distribution.

\begin{figure}
\psfig{figure=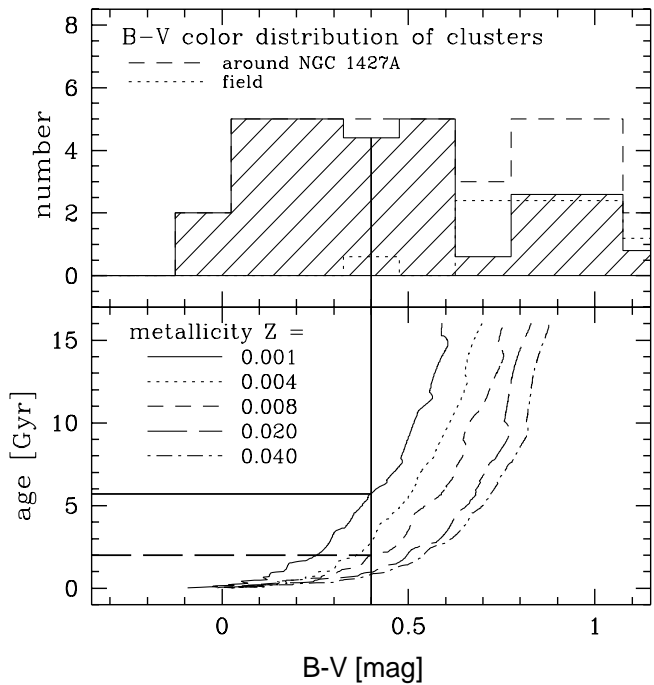,height=8.9cm,width=8.7cm
,bbllx=13mm,bblly=64mm,bburx=125mm,bbury=175mm}
\caption{The hashed histogram shows the $(B-V)$ color distribution of clusters
in
NGC\,1427A after a statistical subtraction of background objects (dotted
histogram). As in Figure 5 the lower panel illustrates the comparison to
color evolution models of clusters.}
\end{figure}

The $(V-I)$ color histogram shows a main concentration of objects around
$(V-I) = 0.7$ mag with a tail to bluer colors and a few objects with
$(V-I) \simeq 0.9$ mag. Colors bluer than $(V-I) = 0.8$ mag can not
be explained by a low metallicity alone, since even the metal poorest 
clusters in our
Milky Way have redder $(V-I)$ colors (e.g. Mc Master catalogue, Harris 1996).
Concerning the $(B-V)$ color most cluster candidates are distributed between
$0.0 < (B-V) < 0.6$ mag. The mean photometric errors are $\sigma_{(V-I)} = 0.08$
mag and $\sigma_{(B-V)} = 0.10$ mag.

\begin{figure}
\psfig{figure=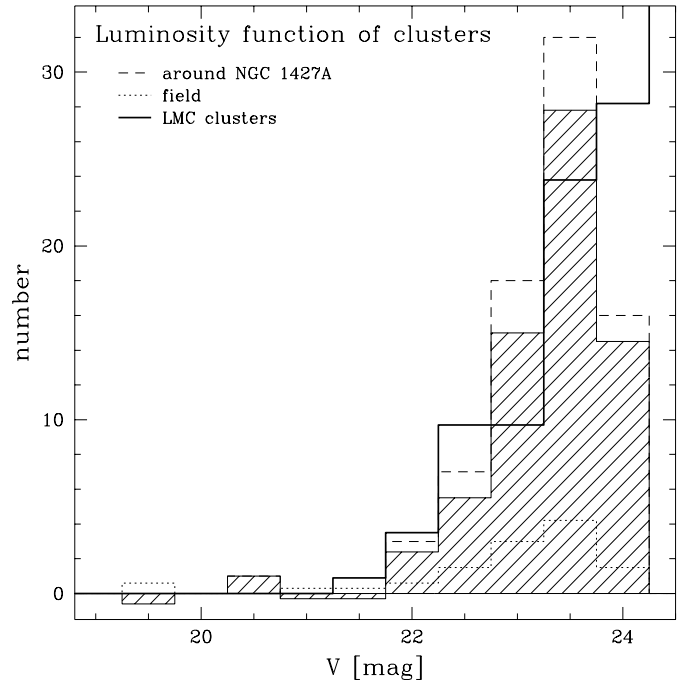,height=8.7cm,width=8.7cm
,bbllx=10mm,bblly=65mm,bburx=195mm,bbury=246mm}
\caption{The hashed histogram shows the $V$ luminosity function of clusters in
NGC\,1427A after a statistical subtraction of background objects (dotted
histogram) from the raw counts (dashed histogram). The brightest objects have
absolute magnitudes of about
$M_V = -9.0$ mag comparable to the luminosity of the brightest LMC clusters.
The thick line represents the counts of the star clusters in the LMC (Bica et
al. 1996) corrected for the Fornax distance and normalized to the luminosity
of NGC\,1427A.}
\end{figure}
     
We can estimate age limits for the objects in dependence of
metallicity. To do so, we used the models of Fritze v.~Alvensleben \& Kurth
(1997, private communication, update of Fritze v.~Alvensleben \& Burkert 1995). 
The lower panels in Figure 5 and 6 illustrate the dependence
between color, age and metallicity for a stellar population that experienced
a single star burst like star clusters.
In the metal poorest ($Z = 0.001$) models, the bulk of the clusters 
with $(V-I) < 0.8$ mag (or $(B-V) < 0.5$ mag) would be younger than 7 Gyr. 
Only the clusters at $(V-I) 
= 1.0$ mag (or $(B-V) = 0.6$ mag) could represent old metal poor 
globular clusters.
Assuming a LMC metallicity ($Z = 0.008$), the reddest clusters
($(V-I) = 0.9$ mag) would have ages of about 5 Gyr, whereas the peak color 
$(V-I) = 0.7$ mag would be consistent with an age of $2\pm 1$ Gyr. This age 
is also 
consistent with the results in the $(B-V)$ color (see Fig.~6).
In this case, the colors as well as the ages of the NGC\,1427A clusters are 
comparable to (but perhaps slightly redder/older than) the intermediate age 
clusters in the LMC (Arimoto \& Bica 1989).

{\small
\begin{table}[b]
\caption{Cluster counts in NGC\,1427A and the LMC}
\begin{tabular}{l r r r}
\hline
 & $M_V < -9.0$ & $M_V < -8.5$ & $M_V < -8.0$\\
\hline
1427A & 0 & 4 & 15\\
LMC & 1 & 11 & 20\\
\hline
\end{tabular}
\end{table}
}

The luminosity function of the cluster candidates appears to be complete down 
to $V\simeq 23$ mag which corresponds to
an absolute magnitude of $M_V\simeq -8$ mag. The brightest objects have 
absolute 
magnitudes of the order of the brightest LMC clusters, as for example 
NGC\,1850 ($M_V = -9$ mag, Bica et al. 1996).
The number of clusters in NGC\,1427A down to $M_V = -8$ mag is about 0.75 times
the cluster counts in the LMC (see Table 2). Normalized to the absolute
luminosity of their host galaxies, the counts are comparable. Figure 7 shows
the luminosity function of the NGC\,1427A clusters in comparison to the one 
of the LMC clusters
after correction for the Fornax distance and normalization to the luminosity
of NGC\,1427A.
The LMC cluster counts are based on
the list of Bica et al. (1996). We included only objects that are classified
as ``star cluster''.
Our analysis shows that NGC 1427A has the richest system observed in an 
irregular galaxy more distant than the Magellanic Clouds up to now.
The old cluster population of the LMC has colors of about 
$(V-I) = 1.0$ mag
and absolute magnitudes $M_V > -8.5$ mag, which is close to our
completeness limit.  The distribution of the old LMC clusters 
is very extended around the LMC, which implies that we will miss 
corresponding objects in NGC 1427A as a consequence of our statistical 
background correction.  Both effects result in a bias against the detection 
of true old clusters in our data.

\section{X-ray properties}

NGC\,1427A is clearly detected with ROSAT PSPC (observation RP600043)
as an X-ray source with 445 $\pm 36$ net photons in the full spectral range of
the ROSAT PSPC (0.1 to 2.4 keV).  The spectrum appears to be relatively
soft with most of the photons at energies lower than 1 keV.  
Neither a fit
with a Raymond-Smith thin plasma, nor a power law gave an acceptably good
fit to the spectrum.
This may imply that the spectrum is a composite of emission from stars,
supernovae, and true diffuse hot gas, typical for an irregular galaxy
(see e.g. Bomans et al. 1997).
We can estimate a total luminosity in the 0.5--2.0 keV band of the ROSAT
PSPC to be 5$ \times 10^{39}$ ergs s$^{-1}$,
adopting our distance modulus and a foreground absorption of
6 $\times 10^{20}$ cm$^{-2}$.
This appears somewhat higher than the only published X-ray luminosity of the
LMC (Wang et al. 1991), but both the Einstein IPC data
of the LMC and the unsatisfactory fit to our ROSAT PSPC spectrum of NGC 1427A
leave a significant margin of uncertainty.
We conclude, that both the X-ray luminosity and hardness of NGC 1427A
are not atypical for a large irregular galaxy (Bomans et al. 1997) and similar
to the values of the LMC.
Much deeper and higher spatial
and spectral resolution data (with the upcoming X-ray satellites AXAF and
XMM) are necessary to investigate the origin of the X-ray emission
of NGC\,1427A further.  Also an integrated spectrum and a total X-ray
luminosity of the LMC derived from the ROSAT All Sky Survey would be of great
value for future studies of the X-ray properties of more distant
star forming galaxies.

\section{Summary}

The irregular galaxy NGC\,1427A is comparable in size, luminosity and color
to the LMC. 
The brightest stellar populations of the galaxy are offset to the 
south-west compared to its nearly elliptical/disky shape at low isophotes. 
At the south-west border we identified a half-ring of OB associations
and H\,{\sc ii} regions, which indicate recent star formation. 
The colors of the `nob' and the low surface brightness tail in 
the east are consistent with starburst ages less than some $10^8$ years,
showing a different star formation history than the main body of NGC\,1427A.
We found about 34 cluster-type objects down to $V = 23.5$ mag that 
are uniformly distributed over the galaxy.
Their color as well as their luminosity distribution is comparable to the 
intermediate age cluster population of the LMC. Assuming a LMC metallicity 
the clusters have mean ages less than 2 Gyr, except a few clusters that might
be genuine old globular clusters.
In the X-ray NGC\,1427A appears similar to the LMC with a relatively 
soft and complex spectrum and a luminosity in the order of 10$^{40}$ ergs 
s$^{-1}$.

NGC\,1427A appears to be nearly a twin of the LMC in all the investigated 
properties. It is interesting to note, that NGC\,1427A shows also 
some signs for tidal influence on the star formation history.  
The ages of clusters point at a dominating cluster formation event a few 
Gyr ago, which accounts for most of the cluster candidates in 
our data.  A sudden event of cluster formation 2 Gyrs ago is well known in the 
LMC (e.g. Bomans et al. 1995, Girardi et al. 1995) and is thought to be 
triggered by interaction with the Milky Way. 
There are several conceivable scenarios that might be responsible for the
present appearence of NGC\,1427A.
Cellone \& Forte (1997) interpreted the `nob' as a separate dwarf galaxy
and concluded that the particular appearence of NGC\,1427A is due to the
collision between two galaxies.
A similar scenario was suggested by Freeman \& de Vaucouleur (1974)
for the folded ring galaxy Arp 144. According to this view the distorted 
appearence and the elongated ring of this galaxy may be due to the encounter
with an intergalactic gas cloud, and possibly triggering starbirth regions
both in the galaxy and the remainder of the cloud. In our case the former
cloud candidate could be the `nob'. Nevertheless, the fact that the `nob'
and the main body of the galaxy are embedded in common and rather symmetrical
envelope may support our interpretation of the `nob' being an original part
of NGC\,1427A.

Alternatively, one may speculate about the passage of NGC\,1427A through the gas
associated with the Fornax cluster as trigger for the subsequent star formation.
If this is the case, the main starburst event must have appeared as the galaxy
started crossing the dense region in the cluster core near NGC 1399 or NGC 1404.
As judged from the cluster candidates in NGC\,1427A, this must have happened 
about 2 Gyr ago.
The question, which of these scenarios should be prefered and whether
the `nob' represents an initially distinct object or belongs to NGC\,1427A, 
can only be solved by dynamical analysis from high
resolution H\,{\sc i} measurements and optical spectra.

\acknowledgements
We thank our referee Dr.~B.E.~Westerlund for helpful comments.
Our paper was partly based on observations made with the NASA/ESA Hubble Space 
Telescope, obtained from the data archive at the Space Telescope Science
Institute. STScI is operated by the Association of Universities for Research
in Astronomy, Inc. under the NASA contract NAS 5-26555.
This research was supported by the {\sc DFG} through the Graduiertenkolleg
`The Magellanic System and other dwarf galaxies' and under 
grant Ri 418/5-1. DJB acknowledge partial support by the Alexander von 
Humboldt-Gesellschaft and thanks S. Points for service observations at
the CTIO 0.9m telescope.
LI wish to thank FONDECYT (grant \# 8970009) and a 1995 Presidential Chair in Science for partial support.


\begin{thebibliography}{}
\bibitem[]{}
Arimoto N., Bica E., 1989, A\&A 222, 89
\bibitem[]{}
Bica E., Clari\'a J.J., Dottori H., Santos J.F.C., Piatti A.E., 1996, ApJS 102,
57
\bibitem[]{}
Bomans, D.J., Chu, Y.-H., Hopp, U. 1997, AJ 113, 1678
\bibitem[]{}
Bomans, D.J., Vallenari A., de~Boer K.S., 1995, A\&A 298, 427
\bibitem[]{}
Bureau M., Mould J.R., Staveley-Smith L., 1996, ApJ 463, 60
\bibitem[]{}
Cellone S.A., Forte J.C., 1997, AJ 113, 1239
\bibitem[]{}
Ferguson H.C., 1989, AJ 98, 367
\bibitem[]{}
Ferguson H.C., Sandage A., 1990, AJ 100, 1
\bibitem[]{}
Freeman K.C., de Vaucouleur G., 1974, ApJ 194, 569
\bibitem[]{}
Fritze-v.~Alvensleben U., Burkert A., 1995, A\&A 300, 58
\bibitem[]{}
Fritze-v.~Alvensleben U., Kurth O., 1997, private communication
\bibitem[]{}
Gallagher J.S., Hunter D.A., 1987, AJ 94, 43
\bibitem[]{}
Girardi L., Chiosi C., Bertelli G., Bressan A., 1995, A\&A 298, 87
\bibitem[]{}
Harris W.E., 1996, AJ 112, 1487
\bibitem[]{}
Huchtmeier W.K., Richter O.-G., 1988, A\&A 203, 237
\bibitem[]{}
Kennicutt R.C., Bresolin F., Bomans D.J., Bothun G.D., Thompson I.B., 1995, AJ
109, 594
\bibitem[]{}
Killeen N.E.B., Bicknell G.V., 1988, ApJ 325,165
\bibitem[]{}
Kissler-Patig M., Kohle S., Hilker M., Richtler T., Infante L., Qintana H., 1996, A\&A 319, 470
\bibitem[]{}
Kohle S., Kissler-Patig M., Hilker M., Richtler T., Infante L., Quintana H.,
1996,  A\&A 309, L39
\bibitem[]{}
Kr\"uger H., Fritze-v.~Alvensleben U., A\&A 284, 793
\bibitem[]{}
Meaburn J., 1980, MNRAS 192, 365
\bibitem[]{}
Roberts M.S., 1975, in "Galaxies and the Universe", eds. A. Sandage,
M. Sandage, and J. Kristian, University of Chicago Press, Chicago, p.~309
\bibitem[]{}
Schombert J.M., 1986, ApJS 60, 603
\bibitem[]{}
Stetson P.B., 1987, PASP 99, 191
\bibitem[]{}
Stetson P.B., 1992, in: "Astronomical Data Analysis Software and Systems I, 
A.S.P. Conference Series, Vol. 25, eds. D.M. Worrall, C. Biemesderfer, and J. 
Barnes, p.~297
\bibitem[]{}
Taylor C.L., Brinks E., Pogge R.W., Skillman, E.D., 1994, AJ 107, 971
\bibitem[]{}
Thuan T.X., 1983, ApJ 268, 667
\bibitem[]{}
Vaucouleurs de G., Vaucouleurs de A., Corwin H.G., Buta R.J.,
Paturel G., Fouqu\'e P., 1991, Third Ref.~Catalogue of Bright
\bibitem[]{}
Westerlund B.E., 1990, A\&AR 2, 29

\end{thebibliography}
\enddocument